\documentclass{article}
\usepackage{IAUS212,graphicx}
\def\edcomment#1{\iffalse\marginpar{\raggedright\sl#1\/}\else\relax\fi}
\reversemarginpar

\begin{document}
\vspace*{1cm}
\title{Resolved and unresolved populations: statistics and synthesis models}
 \author{M. Cervi\~no}
\affil{LAEFF (INTA), Apdo. 50727, 28080 Madrid, Spain}
\affil{IAA (CSIC), Camino Bajo de Hu\'etor 24, 18080 Granada, Spain}

\begin{abstract}
In this contribution, I present the 90\% confidence limits on the
diagnostic diagrams of EW(WR bump) and L(WR bump)/L(H$\beta$) ratio
vs. EW(H$\beta$) resulting from evolutionary synthesis models that 
include the statistical dispersion due to finite stellar populations in 
real star forming regions. 
\end{abstract}

\section{Introduction}

Evolutionary synthesis models have been traditionally used to study the
physical properties of unresolved populations, but they should also be able
to reproduce the integrated properties of resolved ones. In both cases the
discreteness of the stellar populations must be taken into account and
model results must be interpreted in a statistical way, i.e. model outputs
are a {\it mean value} of a probability distribution with an {\it intrinsic
dispersion}. Such dispersion must be taken into account in the
interpretation of the data and it is especially important when the number
of observed stars is small (where the definition of ``small'' depends on
the observable), or in the analysis of surveys.

Especially relevant is the use of diagnostic diagrams, that are, virtually,
independent on the mass/number of stars in the system.  I show in Fig 1 the
90\% confidence limits (CL) for the EW(WR bump) and L(WR bump)/L(H$\beta$)
ratios vs. EW(H$\beta$) for different amount of mass transformed into stars
for evolutionary tracks with standard mass-loss rates and different
metallicities\footnote{Dispersion data values taken from Cervi\~no et
al. (2002), available at {\tt
http://www.laeff.esa.es/users/mcs/SED}.}. 
The plots show the 90\% CL in the results arising from the dispersion in
{\it both axes} and the  correlation coefficients between them; the (Gaussian)
covariance ellipses correspond to an effective number of
stars larger than 10, ${cal N}> 10$, and the rectangular boxes delimit the
uncertainty region for  ${\cal N}< 10$.

\begin{figure}
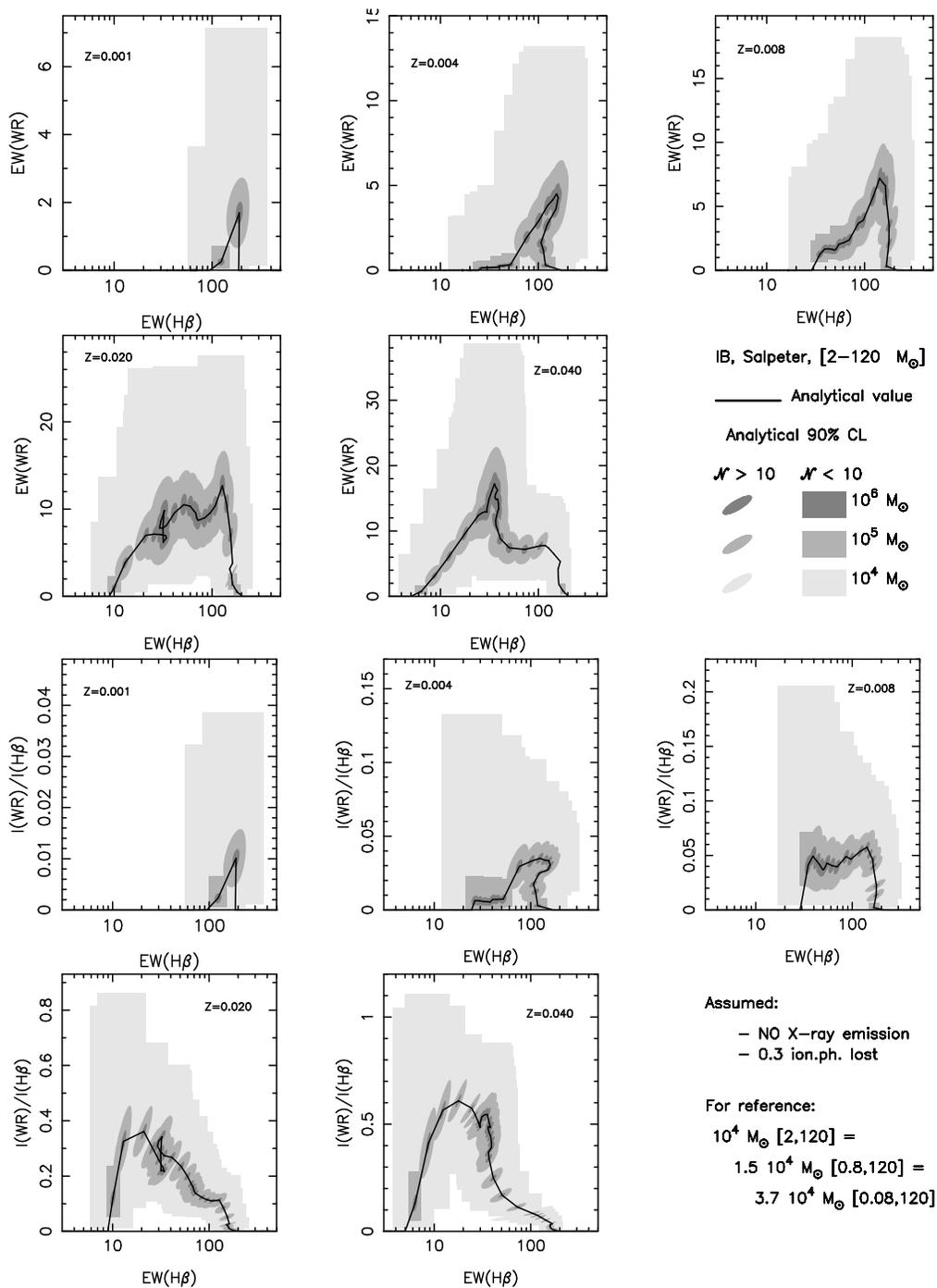

\centering \includegraphics[width=\textwidth]{cerfig1.eps}\\
\centering \includegraphics[width=\textwidth]{cerfig2.eps}
\caption{EW(WR bump) vs.  EW(H$\beta$) and  L(WR bump)/L(H$\beta$) ratio
vs. EW(H$\beta$) for different metallicities.}
\end{figure}


\begin{references}
\reference{Cervi\~no, M., Valls-Gabaud, D., Luridiana, V., \& Mas-Hesse,
J.M. 2002, \aap 381, 51}
\end{references}
\end{document}